\documentclass[lettersize,journal]{IEEEtran}
\usepackage{amsmath,amsfonts}
\usepackage{algorithmic}
\usepackage{array}
\usepackage[caption=false,font=normalsize,labelfont=sf,textfont=sf]{subfig}
\usepackage{textcomp}
\usepackage{url}
\usepackage{verbatim}
\usepackage{graphicx}
\usepackage{booktabs}
\usepackage{multirow}
\usepackage{pifont}    
\usepackage{xcolor}
\usepackage{booktabs}
\usepackage{tabularx}
\newcommand{\redx}{\textcolor{red}{\ding{55}}}
\newcommand{\greencheck}{\textcolor{green!60!black}{\ding{51}}}
\def\BibTeX{{\rm B\kern-.05em{\sc i\kern-.025em b}\kern-.08em
    T\kern-.1667em\lower.7ex\hbox{E}\kern-.125emX}}
\usepackage{balance}
\begin{document}
\title{AffectSpeech: A Large-Scale Emotional Speech Dataset with Fine-Grained Textual Descriptions for Speech Emotion Captioning and Synthesis}
\author{
	Tianhua Qi, \textit{Student Member, IEEE,}
	Wenming Zheng, \textit{Senior Member, IEEE,}
	Bj{\"o}rn W. Schuller, \textit{Fellow, IEEE,}
    Zhaojie Luo, \textit{Member, IEEE,}
    Haizhou Li, \textit{Fellow, IEEE}
\thanks{
	\textit{Corresponding author: Wenming Zheng}.
}
\thanks{Tianhua Qi, Wenming Zheng are with the School of Biological Science and Medical Engineering, Southeast University, Nanjing 210096, China, and also with the Key Laboratory of Child Development and Learning Science (Southeast University), Ministry of Education, China (e-mail: qitianhua@seu.edu.cn, wenming\_zheng@seu.edu.cn).}
\thanks{Bj{\"o}rn W. Schuller is with the Chair of Health Informatics (CHI), Technical University of Munich, 80333 Munich, Germany, and also with the Group on Language, Audio, \& Music (GLAM), Imperial College London, SW7 2AZ London, U.K. (e-mail: schuller@tum.de).}
\thanks{Zhaojie Luo is with the School of Biological Science and Medical Engineering, Southeast University, Nanjing 210096, China, and also with the Shenzhen Loop Area Institute, Shenzhen 518045, China (e-mail: luozhaojie@seu.edu.cn).}
\thanks{Haizhou Li is with the Shenzhen Research Institute of Big Data, School of Artificial Intelligence, The Chinese University of Hong Kong, Shenzhen 518172, China, and also with the Shenzhen Loop Area Institute, Shenzhen 518045, China (e-mail: haizhouli@cuhk.edu.cn).}
}

\markboth{Journal of \LaTeX\ Class Files,~Vol.~18, No.~9, September~2020}%
{How to Use the IEEEtran \LaTeX \ Templates}

\maketitle

\begin{abstract}
Emotion is essential in spoken communication, yet most existing frameworks in speech emotion modeling rely on predefined categories or low-dimensional continuous attributes, which offer limited expressive capacity.
Recent advances in speech emotion captioning and synthesis have shown that textual  descriptions provide a more flexible and interpretable alternative for representing affective characteristics in speech.
However, progress in this direction is hindered by the lack of an emotional speech dataset aligned with reliable and fine-grained natural language annotations.
To tackle this, we introduce AffectSpeech, a large-scale corpus of human-recorded speech enriched with structured descriptions for fine-grained emotion analysis and generation.
Each utterance is characterized across six complementary dimensions, including sentiment polarity, open-vocabulary emotion captions, intensity level,  prosodic attributes, prominent segments, and semantic content, enabling multi-granular modeling of vocal expression.
To balance annotation quality and scalability, we adopt a human–LLM collaborative annotation pipeline that integrates algorithmic pre-labeling, multi-LLM description generation, and human-in-the-loop verification.
Furthermore, these annotations are reformulated into diverse descriptive styles to enhance linguistic diversity and reduce stylistic bias in downstream modeling.
Experimental results on speech emotion captioning and synthesis demonstrate that models trained on AffectSpeech consistently achieve superior performance across multiple evaluation settings.
\end{abstract}


\section{Introduction}
\IEEEPARstart{E}{motion} constitutes an indispensable dimension of spoken communication, encapsulating the speaker’s internal states, underlying  intentions, and interpersonal attitudes that transcend literal lexical meaning~\cite{cowie2003describing}.
As a core component of paralinguistic information, it dictates how messages are perceived and interpreted in social contexts~\cite{van2011emotion}.
Fueled by recent breakthroughs in deep learning and large-scale foundation models~\cite{zhou2025comprehensive}, the fields of speech emotion recognition (SER)~\cite{latif2021survey} and emotional speech synthesis (ESS)~\cite{triantafyllopoulos2023overview} have witnessed substantial progress. These advances have enabled machines to perceive affective cues and generate expressive speech, facilitating a wide range of applications such as empathetic conversational agents~\cite{saffaryazdi2025empathetic} and affect-aware human–computer interaction (HCI) systems~\cite{olugbade2023touch}.

Despite this progress, prevailing paradigms in SER and ESS predominantly rely on categorical labels~\cite{Ekman1999-EKMBE} (e.g., \textit{angry}, \textit{happy}, \textit{sad}) or low-dimensional continuous variables~\cite{EKMAN1955279,russell1980circumplex} (e.g., \textit{valence}, \textit{arousal}, \textit{dominance}).
While computationally convenient, such representations constrain emotional expression to a fixed feature space, struggle to capture the intrinsic complexity, subtlety, and multidimensional nature of human emotions~\cite{lindquist2008emotional}.

To address these limitations, 
recent works~\cite{liang2024aligncap,zhu2024unistyle} have increasingly turned to natural language descriptions as a more expressive and flexible medium for representing speech emotion.
Free-form textual descriptions are capable of articulating nuanced affective states, prosodic characteristics, and communicative intent in ways that transcend predefined taxonomies.
This paradigm shift has given rise to emerging tasks such as speech emotion captioning (SEC)~\cite{xu2024secap}, which aims to generate descriptive sentences characterizing the affective properties of speech, as well as controllable ESS with text-based instructions~\cite{guo2023prompttts}, 
leveraging natural language prompts to guide emotionally nuanced speech generation.

However, progress in these directions is currently constrained by existing emotional speech datasets.
Most available corpora provide only utterance-level annotations, focusing on open-vocabulary descriptions of perceived emotions and a set of coarse prosodic attributes.
For instance, descriptions such as ``\textit{Speaking slowly and with high energy, the woman conveyed her joyful passion effectively}" capture the overall affective impression, yet overlook critical nuances including variations in emotional intensity, salient intervals where emotional expression is most prominent at the segment level, and the intricate interplay between semantic content and prosody at the word level.
Furthermore, due to the prohibitive cost of manual annotation at scale, many corpora~\cite{3754829,10445879} rely heavily on large language models (LLMs) to generate descriptions without systematic human validation, raising concerns regarding annotation reliability and the risk of LLM hallucination.

To overcome these challenges, we introduce \textit{AffectSpeech}, a large-scale emotional speech dataset enriched with fine-grained textual  descriptions to support advanced research in SEC and controllable ESS.
To achieve scalability without compromising annotation quality, we develop a multi-stage human–LLM collaborative pipeline that integrates algorithmic  pre-labeling, multi-LLM description generation, and human-in-the-loop verification, effectively balanceing annotation cost, linguistic richness, and data reliability.
Each utterance is annotated across six complementary dimensions to support multi-granular emotional analysis:
\begin{enumerate}
	\item \textbf{Sentiment polarity}: Categorizing utterances into positive, negative, or neutral valences to provide a coarse-grained directional anchor.
	\item \textbf{Open-vocabulary emotion captions}: Providing free-form descriptions that articulate nuanced and idiosyncratic emotional states beyond predefined taxonomies.
	\item \textbf{Emotion intensity}: Quantifying affective strength to model dynamic fluctuations of emotional states.
	\item \textbf{Prosodic descriptors}: Characterizing pitch, energy, and tempo to link acoustic variations to underlying intent.
	\item \textbf{Prominent segments}: Identifying temporal intervals where emotional expression is most salient.
	\item \textbf{Semantic content}: Capturing emotion-related keywords to facilitate context-aware understanding.
\end{enumerate}

Moreover, drawing on Systemic Functional Linguistics~\cite{eggins2004introduction}, which posits that language fundamentally varies according to its situational context, we transform these annotations into six functional styles: \textit{narrative}, \textit{profiling}, \textit{synopsis}, \textit{bullet-point}, \textit{technical}, and \textit{structural}.
This diversification prevents speech language models (SpeechLMs) from overfitting to rigid phrasing templates and enhances generalization, thereby better addressing diverse user demands.

The main contributions of this paper are listed as follows:
\begin{itemize}
	\item We introduce \textit{AffectSpeech}, the first large-scale emotional speech dataset with fine-grained, multi-level textual descriptions, enabling advanced research in SEC and ESS.
	\item We develop a scalable human–LLM collaborative annotation pipeline that integrates algorithmic pre-labeling, multi-LLM description generation, and human-in-the-loop verification, ensuring the  reliability, validity, and linguistic diversity of the resulting descriptions.
	\item Extensive experiments demonstrate that \textit{AffectSpeech} 
    considerably 
    enhances models' performance in interpretable emotion analysis and controllable emotional speech generation across multiple downstream tasks.
\end{itemize}

Dataset and demo samples are available at  \url{https://github.com/jeremychee4/AffectSpeech}.

\section{Background and Related Work}
\subsection{Emotional Speech Dataset}
The advancement of emotional speech research has been closely tied to the availability of high-quality datasets. Based on their annotation paradigms, existing emotional speech corpora can be broadly categorized into three groups: 1) categorical, 2) dimensional, and 3) descriptive. A comparison of representative datasets is summarized in Table~\ref{tab:emotion_datasets}.
\subsubsection{Categorical emotional speech datasets}
The categorical paradigm, largely inspired by Ekman’s theory of basic emotions~\cite{Ekman1999-EKMBE}, remains the most widely adopted annotation scheme in speech emotion research. Early datasets such as EMO-DB~\cite{burkhardt05b_interspeech}, SAVEE~\cite{jackson2014surrey}, and RAVDESS~\cite{livingstone2018ryerson} provide carefully controlled acted speech annotated with discrete emotion labels (e.g., angry, happy, sad), and have been extensively used for benchmarking SER systems~\cite{liu2024improving}.
Despite their popularity and ease of use, categorical datasets inherently impose a mutually exclusive view of affective states~\cite{russell1980circumplex}. Such representations are limited in their ability to capture mixed emotions, subtle expressive variations, or gradual emotional transitions~\cite{larsen2014case,frijda1992complexity}, all of which are critical for fine-grained emotion understanding and expressive speech synthesis.
\subsubsection{Dimensional emotional speech datasets}
To address the limitations of predefined categories, dimensional datasets represent emotional states in a continuous space 
following the early works of Wundt~\cite{EKMAN1955279,russell1980circumplex}, most commonly using valence, arousal, and dominance. Representative corpora such as IEMOCAP~\cite{busso2008iemocap}, MSP-IMPROV~\cite{Busso_2017}, and MSP-Podcast~\cite{lotfian2017building} provide numerical annotations that enable the modeling of emotional dynamics and intensity variations across samples~\cite{qi24_interspeech}.
While dimensional representations offer improved expressive resolution compared to categorical labels~\cite{Ekman1999-EKMBE}, they can be abstract and unintuitive for direct human interaction in many application scenarios~\cite{Russell2003CoreAA}. In particular, continuous affective coordinates are challenging for users to interpret and specify as control signals~\cite{BRADLEY199449}, which limits their applicability in interactive systems especially for controllable emotional speech synthesis.
\subsubsection{Descriptive emotional speech datasets}
Since human language is the most intuitive medium to express emotional intent~\cite{de1981introduction}, recent research~\cite{101145,lengprompttts} has increasingly explored descriptive emotional speech datasets that leverage free-form text to characterize affective and prosodic properties.
For instance, EmoVoice-DB~\cite{3754829} introduces natural language  descriptions aligned with emotional speech. However, it is composed entirely of TTS-synthesized audios and remains limited in scale. This reliance on synthetic speech may constrain the acoustic variability and emotional authenticity, which are essential for SEC and ESS.
TextrolSpeech~\cite{10445879} scales up authentic speeches and incorporates both prosodic and emotional descriptions. Nevertheless, its annotations are entirely generated by LLMs at the utterance level, focusing on  coarse-grained summaries. The absence of human verification may raise concerns regarding annotation reliability. 
In contrast, \textit{AffectSpeech} bridges these gaps by providing a large-scale corpus of authentic, human-recorded speech paired with multi-granular, human-validated natural language annotations.
\begin{table*}[htp]
	\centering
	\footnotesize
	\setlength{\tabcolsep}{5pt}
	\caption{Comparison of representative emotional speech datasets based on their provided annotation types and data characteristics.}
	\label{tab:emotion_datasets}
	
	\begin{tabular}{
			l r c l l
			c c c c c c
			r l
		}
		\toprule
		\textbf{Dataset} &
		\textbf{\# Samples} &
		\textbf{\# Emo.} &
		\textbf{Elicitation} &
		\textbf{Desc. Form} &
		\textbf{Sent.} &
		\textbf{Emo*} &
		\textbf{Pros.} &
		\textbf{Int.} &
		\textbf{Seg.} &
		\textbf{Sem.} &
		\textbf{\# Desc.} &
		\textbf{Anno. by} \\
		\midrule
		
		\multicolumn{13}{l}{\textit{Categorical}} \\
		\addlinespace[2pt]
		SAVEE~\cite{jackson2014surrey}        & 480      & 7 & Acted      & Label & \redx & \redx & \redx & \redx & \redx & \redx & 480      & Human \\
		EMO-DB~\cite{burkhardt05b_interspeech}        & 535      & 7 & Acted      & Label & \redx & \redx & \redx & \redx & \redx & \redx & 535      & Human \\
		eNTERFACE~\cite{1623803}    & 1,263    & 6 & Acted      & Label & \redx & \redx & \redx & \redx & \redx & \redx & 1,263    & Human \\
		JL-Corpus~\cite{james18_interspeech}         & 2,400& 5 & Acted    & Label & \redx & \redx & \redx & \redx & \redx & \redx & 2,400 & Human \\
		TESS~\cite{E8H2MF_2020}         & 2,800    & 7 & Acted      & Label & \redx & \redx & \redx & \redx & \redx & \redx & 2,800    & Human \\
		CREMA-D~\cite{cao2014crema}      & 7,442    & 6 & Acted      & Label & \redx & \redx & \redx & \redx & \redx & \redx & 7,442    & Human \\
		EmoV-DB~\cite{adigwe2018emotional}         & 7,590& 5 & Acted    & Label & \redx & \redx & \redx & \redx & \redx & \redx & 7,590 & Human \\
		ESD~\cite{zhou2022emotional}          & 35,000   & 5 & Acted      & Label & \redx & \redx & \redx & \redx & \redx & \redx & 35,000   & Human \\
		MELD~\cite{poria2019meld}         & 13,706   & 7 & Spontaneous     & Label & \greencheck & \redx & \redx & \redx & \redx & \redx & 13,706 & Human \\
		RAVDESS~\cite{livingstone2018ryerson}      & 7,356    & 8 & Acted      & Label & \redx & \redx & \redx & \greencheck & \redx & \redx & 7,356 & Human \\
		MEAD~\cite{kaisiyuan2020mead}         & 32,160& 8 & Induced    & Label & \redx & \redx & \redx & \greencheck & \redx & \redx & 32,160 & Human \\
		
		\midrule
		\multicolumn{13}{l}{\textit{Dimensional}} \\
		\addlinespace[2pt]
		MESS~\cite{morgan2019categorical}         & 1,800    & 4 & Acted   & Numeric & \multicolumn{6}{l}{\textit{Valence, Arousal}} & 1,800 & Human \\
		MSP-IMPROV~\cite{Busso_2017}   & 7,818    & 4 & Hybrid  & Numeric & \multicolumn{6}{l}{\textit{Valence, Arousal, Dominance}} & 7,818 & Human \\
		IEMOCAP~\cite{busso2008iemocap}      & 10,039   & 5 & Hybrid  & Numeric & \multicolumn{6}{l}{\textit{Valence, Arousal, Dominance}} & 10,039 & Human \\
		MSP-Podcast~\cite{lotfian2017building}  & 264,705  & 8 & Spontaneous  & Numeric & \multicolumn{6}{l}{\textit{Valence, Arousal, Dominance}} & 264,705 & Human \\
		
		\midrule
		\multicolumn{13}{l}{\textit{Descriptive}} \\
		\addlinespace[2pt]
		EmoVoice-DB~\cite{3754829}   & 22,100   & 7 & Synthetic & Text & \redx & \greencheck & \redx & \redx & \redx & \redx & 64,200 & LLM \\
		TextrolSpeech~\cite{10445879} & 236,203  & 8 & Hybrid    & Text & \redx & \greencheck & \greencheck & \redx & \redx & \redx & 236,203 & LLM \\
		\textbf{AffectSpeech} 
		& 253,799       & 9 & Hybrid    & Text 
		& \greencheck & \greencheck & \greencheck & \greencheck & \greencheck & \greencheck 
		& 1,522,794 & LLM + Human \\
		
		\bottomrule
	\end{tabular}
	
	\vspace{1mm}
	\raggedright
	\footnotesize
	\textit{Notes:}
	\# Emo. = Number of basic emotion categories;
	Desc. Form = Description form;
	Sent. = Sentiment polarity;
	Emo* = Open-vocabulary emotion captions;
	Pros. = Prosody;
	Int. = Intensity;
	Seg. = Prominent segments;
	Sem. = Semantic content analysis;
	\# Desc. = Number of descriptions/labels;
	Anno. by = Source of annotation.
	Only annotations that are explicitly defined and publicly available in the original datasets are considered in this comparison.
\end{table*}
\subsection{Speech Emotion Captioning}
Speech emotion captioning~\cite{xu2024secap} has emerged as a promising research direction at the intersection of audio-to-text generation and affective computing. Unlike traditional SER~\cite{latif2021survey} which predicts predefined categories or attributes, SEC aims to generate semantically rich and descriptive text that characterizes the emotional and prosodic nuances of speech.

Previous frameworks~\cite{liang2024aligncap,10447102} attempted to bridge the gap between acoustic features and the semantic space through cross-modal alignment techniques such as contrastive language-audio pretraining (CLAP)~\cite{elizalde2023clap} and Q-Former~\cite{li2023blip}.
More recently, the field has shifted towards leveraging large-scale foundation models via supervised fine-tuning (SFT) to further enhance captioning performance.
For instance, advanced models such as Qwen2-Audio~\cite{chu2024qwen2} and Qwen2.5-Omni~\cite{xu2025qwen2} have been fine-tuned to produce expressive and fluid descriptions by inheriting robust pre-trained multimodal alignments.
Despite their impressive interactive capabilities, these models still struggle with in-depth emotional analysis.
In practice, they exhibit significant limitations in quantifying emotional intensity and identifying prominent segments within an utterance. Furthermore, these models often fail to link prosodic attributes, such as pitch excursions and rhythmic shifts, with the underlying emotional state, leading to descriptions that lack granular precision.
By performing SFT on \textit{AffectSpeech}, these foundation models are expected to learn the intricate mappings between acoustic realizations and multi-granular textual descriptions, thereby enabling more accurate, nuanced, and interpretable SEC.
\subsection{Emotional Speech Synthesis with Textual Description}
\begin{figure*}[ht]
	\centering
	\includegraphics[width=2.0\columnwidth]{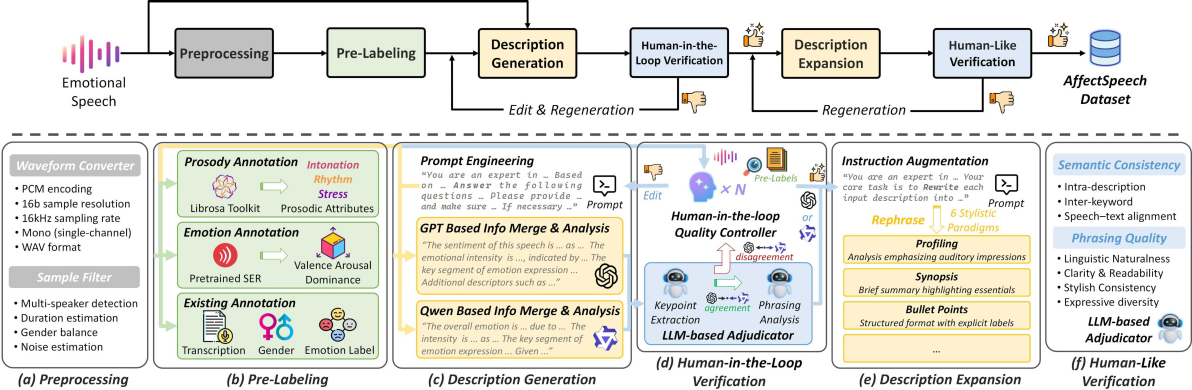}
	\caption{The construction and annotation pipeline of the AffectSpeech dataset.}
	\label{fig-framework}
\end{figure*}
Emotional speech synthesis~\cite{triantafyllopoulos2023overview} aims to generate speech that conveys not only linguistic content but also controllable and perceptually appropriate emotions. This is typically achieved through prosodic modulations of intonation, rhythm, and intensity~\cite{cole2015prosody}.
Conventional emotional TTS~\cite{20223145293,11049047} and voice conversion (VC)~\cite{10446191,10889084} primarily rely on discrete emotion labels or reference embeddings extracted from reference speech.
To achieve more flexible and interpretable control, recent studies~\cite{lengprompttts,qi25_interspeech} have shifted toward description-based ESS, where natural language prompts serve as the primary conditioning information.

Early works~\cite{guo2023prompttts,yang2024instructtts} demonstrated the feasibility of using textual descriptions to manipulate emotional expression.
Built upon this, large-scale generative models such as  VoxInstruct~\cite{zhou2024voxinstruct} have further explored in-context learning paradigms that incorporate descriptive human instruction for emotion synthesis.
Despite these advancements, existing methodologies are primarily limited to processing utterance-level instructions, which lack the ability to provide feedback on localized affective details. Consequently, these systems struggle to respond to nuanced instructions that require precise control over localized emotional outbreaks, fine-grained intensity transitions, or the subtle alignment of prosodic emphasis with specific lexical content.
By performing SFT on \textit{AffectSpeech}, which offers diverse, fine-grained, and human-validated descriptions, ESS systems can learn to bridge the gap between high-level instructions and low-level prosodic realization, ultimately enabling more expressive and controllable emotional speech synthesis.

\section{AffectSpeech Dataset}
\subsection{Overview of AffectSpeech}
The proposed \textit{AffectSpeech} is a large-scale emotional speech dataset enriched with fine-grained textual descriptions to facilitate alignment between high-fidelity speech waveforms and emotion captions. The dataset contains 253,799 authentic English utterances produced by human speakers, covering nine basic emotion categories (\textit{angry}, \textit{disgust}, \textit{fear}, \textit{happy}, \textit{neutral}, \textit{sad}, \textit{surprise}, \textit{contempt}, \textit{calm}),  alongside a broad spectrum of expressive speaking styles. Each utterance is annotated with multi-aspect natural language descriptions that comprehensively characterize emotional states, prosodic attributes, intensity levels, prominent segments, and semantic cues.
In total, \textit{AffectSpeech} provides 1,522,794 textual descriptions generated via a collaborative LLM–human annotation framework. The data construction pipeline is illustrated in Fig.~\ref{fig-framework}.
\subsection{Data Collection and Preprocessing}
\begin{figure}[btp]
	\centering
	\includegraphics[width=1.0\columnwidth]{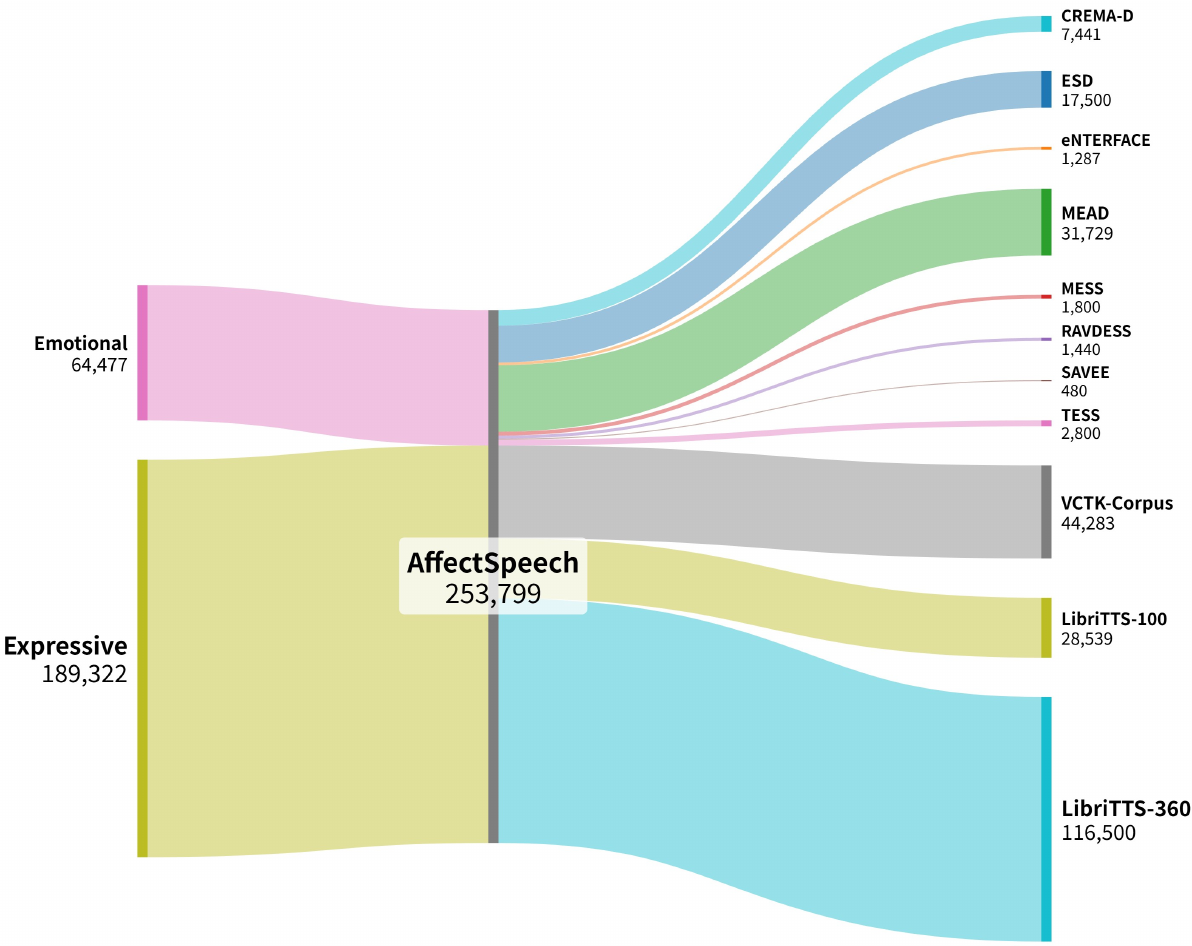}
	\caption{Composition and scale of the AffectSpeech dataset categorized by original data sources.}
	\label{fig-data-source}
\end{figure}
The construction of \textit{AffectSpeech} follows a set of principled criteria to ensure emotional expressiveness, acoustic quality, and broad research utility. We prioritize recordings with balanced gender distribution, low background noise, and high fidelity, with a preference for single-speaker utterances. To promote reproducibility and legal compliance, we exclusively aggregate publicly available datasets with clear licensing terms.
Based on these criteria, \textit{AffectSpeech} integrates utterances from SAVEE~\cite{jackson2014surrey}, RAVDESS~\cite{livingstone2018ryerson}, eNTERFACE~\cite{1623803}, TESS~\cite{E8H2MF_2020}, CREMA-D~\cite{cao2014crema}, ESD~\cite{zhou2022emotional},  MEAD~\cite{kaisiyuan2020mead}, and MESS~\cite{morgan2019categorical}. To further enhance lexical diversity and natural speaking styles, we incorporate subsets from the VCTK-Corpus~\cite{yamagishi2019cstr} and  LibriTTS~\cite{zen19_interspeech}. The composition and scale of the resulting corpus are summarized in Fig.~\ref{fig-data-source}.

All recordings undergo a preprocessing pipeline to ensure cross-source consistency. Audio signals are standardized to a standard WAV format with PCM encoding, 16-bit resolution, a 16~kHz sampling rate, and mono-channel configuration. Sample-level filtering is then applied, including multi-speaker detection, duration estimation, and signal-to-noise ratio (SNR) assessment. These procedures establish a unified foundation for downstream annotation.
\begin{figure*}[tbp]
	\centering
	\includegraphics[width=1.98\columnwidth]{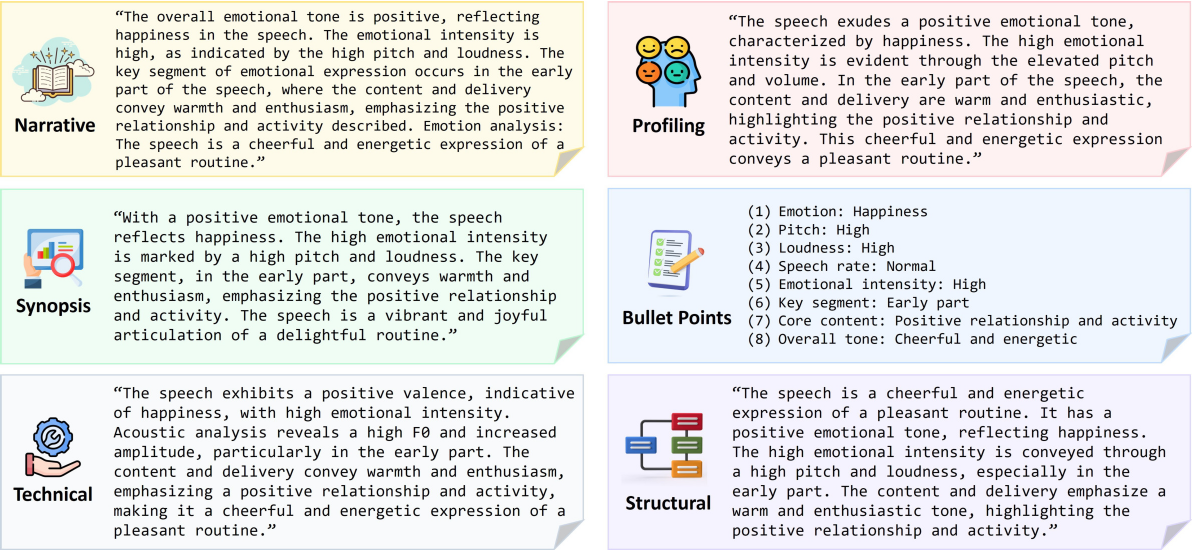}
	\caption{Illustration of diverse speech emotion captioning styles.}
	\label{fig-desc-demo}
\end{figure*}
\subsection{Pre-Labeling}
Prior to natural language annotation, an algorithmic pre-labeling stage is performed to extract structured affective priors from speech waveforms and existing metadata. These automatically derived attributes serve as auxiliary variables for subsequent LLM-based description generation and human-in-the-loop verification, thereby enhancing annotation consistency and interpretability.
\subsubsection{Prosody annotation}
Prosodic attributes are extracted using the Librosa toolkit\footnote{\url{https://librosa.org/}}~\cite{mcfee2015librosa} to characterize the acoustic realization of emotional expression. Specifically, pitch, speaking rate, and short-time energy are utilized to model intonation patterns, rhythmic variations, and loudness/stress, respectively.
\subsubsection{Emotion annotation}
A pretrained SER  model\footnote{\url{https://github.com/audeering/w2v2-how-to}}~\cite{wagner2023dawn} is employed to predict valence–arousal–dominance at the utterance level. These continuous estimates complement discrete emotion categories by capturing emotional intensity and subtle affective variations.
\subsubsection{Existing annotations}
We further leverage existing annotations from the source datasets, including transcriptions, speaker gender, and predefined emotion labels. These metadata provide essential semantic grounding and serve as references for quality control throughout the annotation process.
\subsection{Description Generation}
To scale the production of fine-grained textual descriptions, we leverage state-of-the-art multimodal LLMs (MLLMs). Each model is conditioned on a structured prompt encompassing speech waveforms, prosodic characteristics,  VAD values, and available metadata including transcriptions, speaker gender, and categorical emotion labels.
Through instruction-based prompting, the MLLMs integrate acoustic and semantic cues to generate open-vocabulary natural language descriptions. The resulting captions characterize sentiment polarity, nuanced emotional states, intensity levels, and prosodic descriptors, alongside prominent segments and semantic content.

To mitigate model-specific biases and improve robustness, we employ a multi-LLM generation strategy incorporating  GPT-based\footnote{\url{https://platform.openai.com/docs/guides/audio}} and Qwen-based\footnote{\url{https://qwen.ai/apiplatform}} MLLMs. This ensemble approach facilitates a comparative evaluation of generation quality, ensuring the selection of highly accurate and expressive descriptions during subsequent verification.
\begin{figure}[tbp]
	\centering
	\includegraphics[width=1.0\columnwidth]{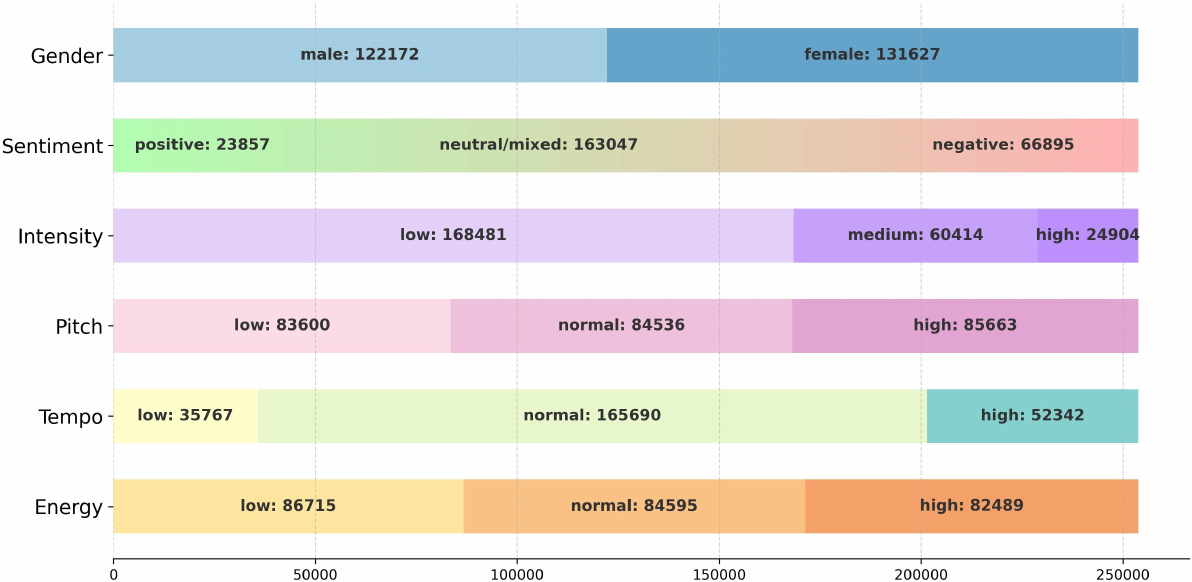}
	\caption{Quantitative distributions of speaker gender, sentiment polarity, intensity level, and prosodic attributes (pitch, tempo, and energy).}
	\label{fig-label-distribution}
\end{figure}
\begin{table}[ht]
	\centering
	\small
	\caption{Statistical distribution of LLM adjudication and human verification outcomes.}
	\label{tab:human_verification}
	\begin{tabular}{lrr}
		\toprule
		\textbf{Verification Stage}  & \textbf{Count} & \textbf{Proportion [\%]} \\
		\midrule
		Direct Consensus\textsuperscript{a} & 164,747 & 64.9 \\
		\midrule
		Manual Selection\textsuperscript{b} & 73,468 & 28.9 \\
		Manual Correction\textsuperscript{c} & 15,584 & 6.2 \\
		\midrule
		\textbf{Total} & \textbf{253,799} & \textbf{100.0} \\
		\bottomrule
		\addlinespace[2pt]
		\multicolumn{3}{l}{\scriptsize \textsuperscript{a} Descriptions from multiple MLLMs exhibited high keypoint consistency.} \\
		\multicolumn{3}{l}{\scriptsize \textsuperscript{b} Human experts identified one valid candidate despite keypoint discrepancies.} \\
		\multicolumn{3}{l}{\scriptsize  \textsuperscript{c} Both candidates failed; involved metadata correction and regeneration.}	
	\end{tabular}
\end{table}
\subsection{Human-in-the-Loop Verification}
To ensure annotation reliability while maximizing scalability, we adopt a human-in-the-loop verification framework augmented with an LLM-based adjudication mechanism. Specifically, an LLM adjudicator first extracts keypoints from the descriptions generated by multiple MLLMs, focusing on core attributes such as sentiment polarity, emotional intensity, prosodic characteristics, and prominent segments.
If the extracted keypoints exhibit high inter-model consistency, the adjudicator performs phrasing analysis to select the candidate with the most coherent structure and highest expressive quality. Conversely, when discrepancies arise, the sample is escalated to a human quality controller. The controller evaluates the audio and associated metadata to determine which description best aligns with the actual emotional expression.
If neither candidate adequately reflects the auditory content, the human annotator performs metadata correction and refines the prompt to trigger a regeneration cycle via the MLLM, thereby closing the verification loop.

The human quality control was conducted by a group of five expert annotators (3 female, 2 male; age $27.4 \pm 3.1$). All participants had academic backgrounds in linguistics or psychology, ensuring professional judgment on prosodic nuances and emotional descriptors. Prior to the task, each annotator completed a 4-hour training session to align on the multi-aspect annotation criteria.
To quantify the impact of human intervention, we tracked the adjudication outcomes across the entire corpus, as detailed in Table~\ref{tab:human_verification}. The results indicate that MLLMs achieved a high autonomous consensus rate of 64.9\%, and human experts were required to  resolve the remaining 35.1\% of samples. Specifically, 28.9\% of the corpus was resolved via expert selection of an existing candidate, while 6.2\% required manual correction and description regeneration. In this context, multiple regeneration attempts for a single sample are consolidated into a single recorded intervention.
\begin{figure}[tbp]
	\centering
	\includegraphics[width=1.0\columnwidth]{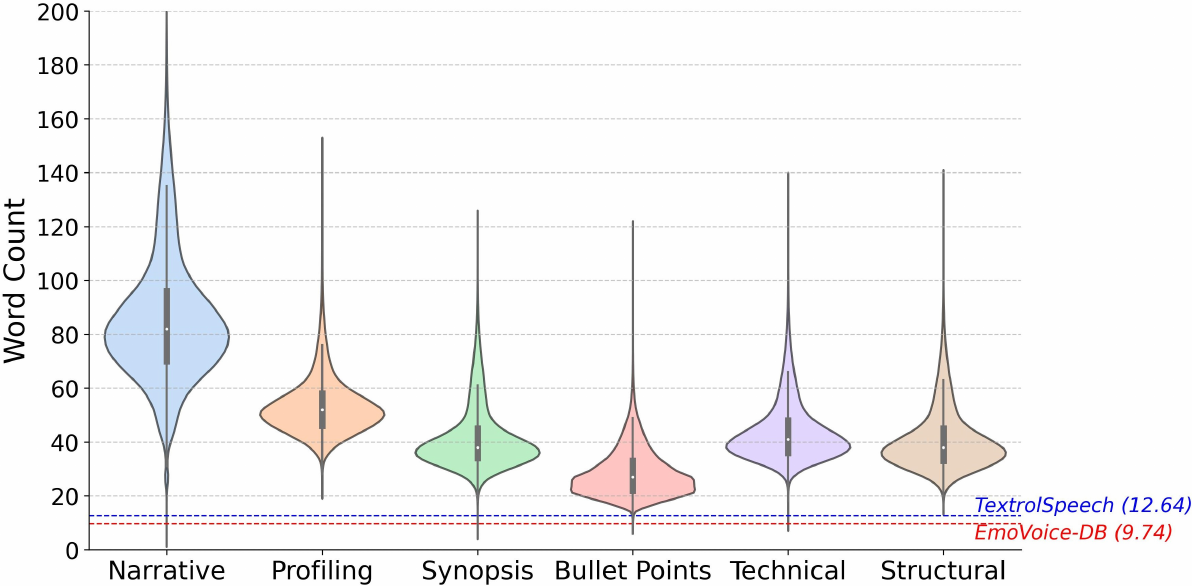}
	\caption{Text length distributions across six textual description styles.}
	\label{fig-desc-distribution}
\end{figure}
\subsection{Description Expansion}
Each verified description is expanded into multiple stylistic variants using LLMs, as illustrated in Fig.~\ref{fig-desc-demo}. This design allows a single utterance to be represented across diverse perspectives. Specifically, six functional styles are generated:
\begin{enumerate}
	\item \textbf{Narrative}: A cohesive, fluent paragraph integrating all descriptive aspects.
	\item \textbf{Profiling}: An analytical description focusing on auditory impressions and subtle emotional nuances.
	\item \textbf{Synopsis}: A concise summary emphasizing important affective and prosodic characteristics.
	\item \textbf{Bullet Points}: A point-wise format with explicit labels (e.g., Emotion: Happiness) to be clearly seen.
	\item \textbf{Technical}: A formal description incorporating technical acoustic indicators, such as \textit{F0} and \textit{amplitude}.
	\item \textbf{Structural}: A hierarchical representation consisting of a core statement followed by supplementary details.
\end{enumerate}
\subsection{Human-Like Verification}
To assess the quality of expanded descriptions at scale and ensure overall reliability, we employ an additional LLM-based adjudicator that approximates human behavior for quality control. This adjudicator performs multi-aspect semantic and linguistic validation, thereby reducing residual annotation noise without incurring manual cost.
Each description is evaluated in terms of intra-description consistency, speech–text alignment, linguistic naturalness, clarity and readability, stylistic adherence, and expressive diversity. Descriptions that fail to meet any criterion are returned for regeneration, while validated outputs are incorporated into the final description pool.
\begin{figure}[tp]
	\centering
	\includegraphics[width=1.0\columnwidth]{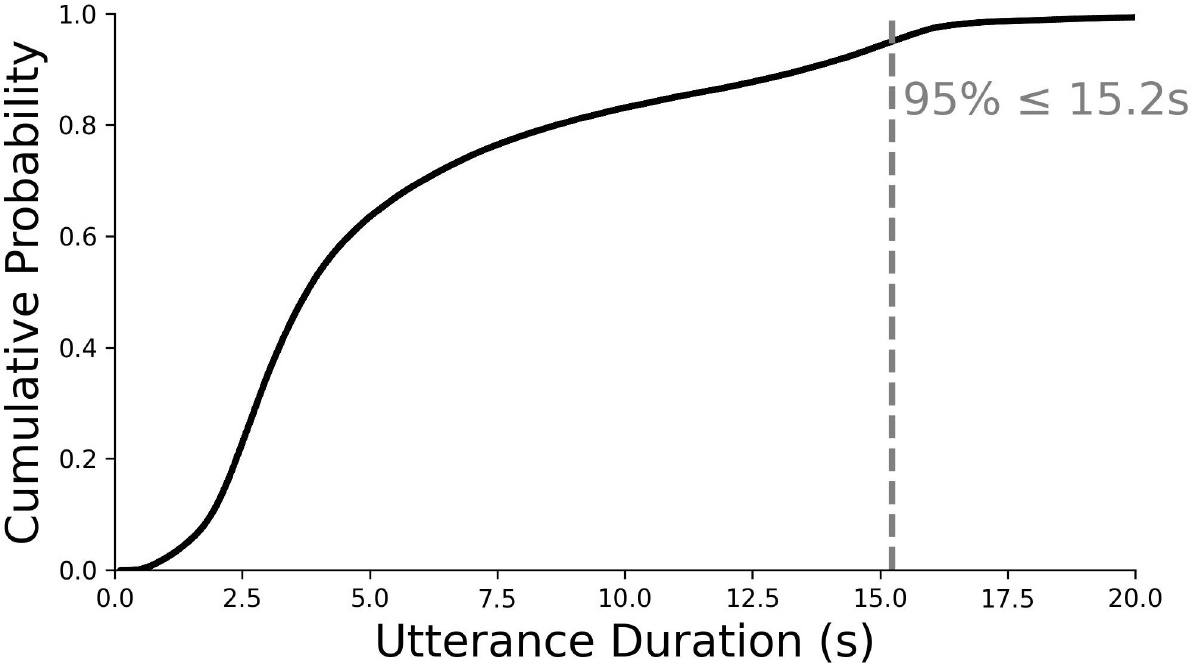}
	\caption{Empirical cumulative distribution function of utterance duration.}
	\label{fig-dur-ecdf}
\end{figure}
\begin{figure}[bp]
	\centering
	\includegraphics[width=1.0\columnwidth]{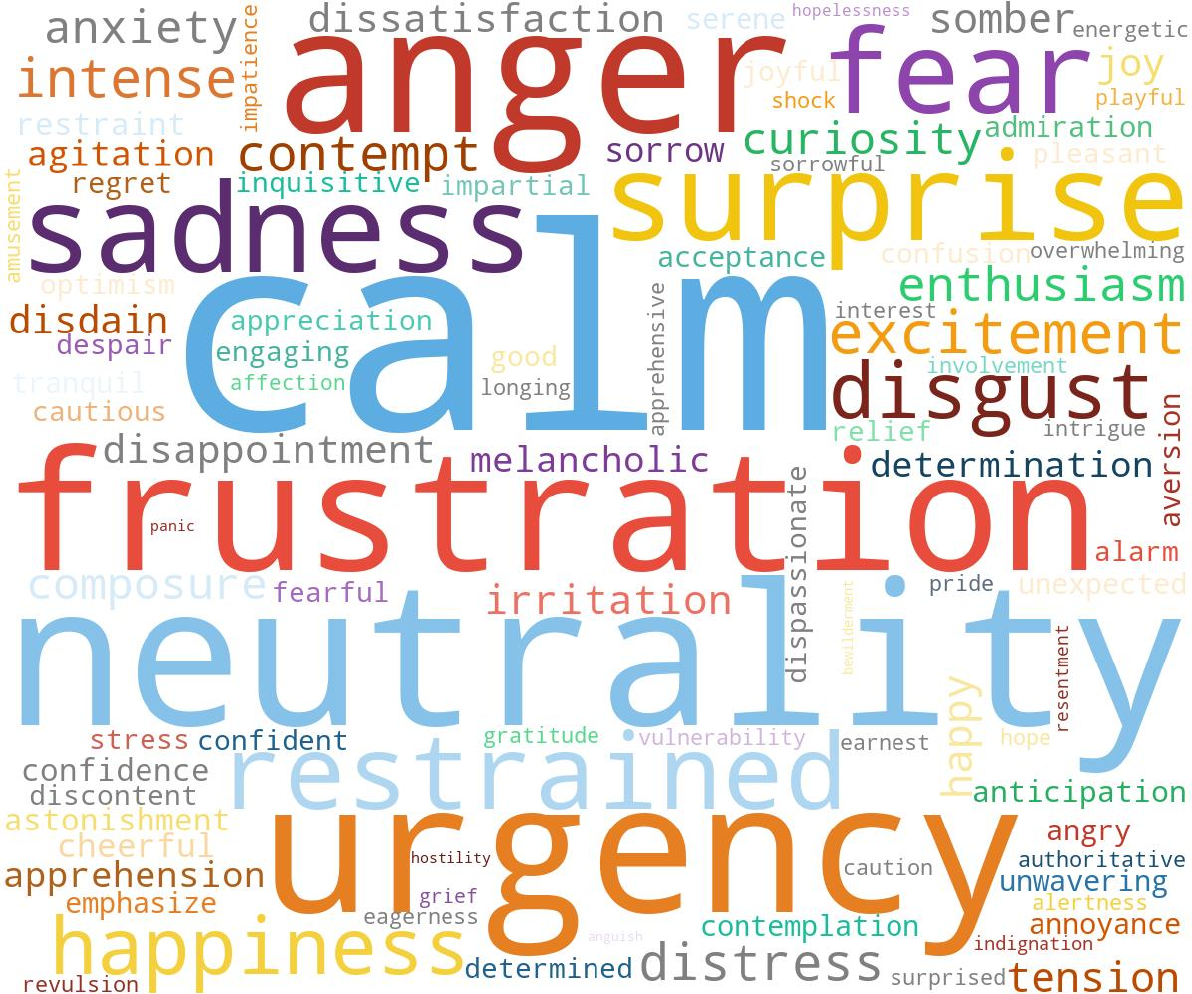}
	\caption{Word cloud of the top-100 most frequent emotion-related words.}
	\label{fig-word-cloud}
\end{figure}
\subsection{Statistics}
This subsection provides a comprehensive statistical analysis of \textit{AffectSpeech}, encompassing acoustic–affective attributes and linguistic properties of the textual descriptions. All statistics are computed over the full dataset and illustrated through multiple complementary visualizations.
\subsubsection{Affective and prosodic distributions}
Fig.~\ref{fig-label-distribution} presents the distributions of speaker gender, sentiment polarity, emotion intensity, pitch, tempo, and energy. The dataset exhibits a well-balanced gender ratio and broad coverage across sentiment polarities and prosodic dimensions. Notably, low-intensity and neutral/mixed emotions dominate, reflecting the prevalence of subtle, and context-dependent emotions in natural speech.
\subsubsection{Text length across description styles}
Fig.~\ref{fig-desc-distribution} compares text length distributions across the six description styles. Narrative descriptions are the most detailed, whereas synopsis and bullet-point formats are considerably more concise, forming a clear hierarchy of linguistic granularity. Compared with existing emotion captioning datasets, \textit{AffectSpeech} offers substantially richer and more diverse textual information.
\subsubsection{Utterance duration}
The empirical cumulative distribution function (ECDF) of utterance duration is shown in Fig.~\ref{fig-dur-ecdf}. Approximately 95\% of the utterances are shorter than 15.2 seconds, indicating a moderate temporal span that is suitable for multi-level emotion modeling.
\subsubsection{Prominent segments}
Emotionally salient intervals are most frequently observed at the beginning of utterances (28.2\%), followed by the medial (24.7\%) and final segments (22.4\%). Additionally, 21.4\% of the samples exhibit a consistent emotional expression throughout the entire utterance, while the remaining 3.3\% are annotated as unclear.
\subsubsection{Emotion lexicon}
Fig.~\ref{fig-word-cloud} visualizes the top 100 emotion-related words in the corpus. The vocabulary encompasses both categorical emotion labels and fine-grained affective modifiers, reflecting the open-vocabulary design and the descriptive richness of the annotations.
\begin{table*}[ht]
	\centering
	\footnotesize
	\setlength{\tabcolsep}{7pt}
	\caption{Performance of Speech Emotion Captioning across different models and supervised Fine-tuning datasets.}
	\label{tab:sec}
	
	\begin{tabular}{
			l
			ccc
			ccc
			ccc
		}
		\toprule
		\multirow{2}{*}{\textbf{Model}} &
		\multicolumn{3}{c}{\textbf{Emotion Accuracy [\%] $\uparrow$}} &
		\multicolumn{3}{c}{\textbf{Prosody Accuracy [\%] $\uparrow$}} &
		\multicolumn{3}{c}{\textbf{Description Quality}} \\
		
		\cmidrule(lr){2-4}
		\cmidrule(lr){5-7}
		\cmidrule(lr){8-10}
		
		& \textbf{Category} & \textbf{Intensity} & \textbf{Key Segment}
		& \textbf{Pitch} & \textbf{Tempo} & \textbf{Energy}
		& \textbf{Distinct-2} & \textbf{Self-BLEU} & \textbf{Avg. Word} \\
		\midrule
		
		GPT-4o mini Audio  & 58.88 & 36.75 & 32.63 & 37.50 & 64.00 & 31.75 & 0.4435 & 0.5936 & 112.81 \\
		GPT-4o Audio       & 66.50 & \textbf{57.13} & 37.25 & 41.75 & 64.25 & 36.38 & 0.3439 & 0.6975 & 101.50 \\
		Step-Audio 2 mini~\cite{wu2025step}  & 66.63 & 45.00 & 35.75 & 33.38 & 64.75 & 38.25 & 0.1867 & 0.8598 & 84.00 \\
		Audio Flamingo 3~\cite{ghosh2025audio}  & 70.50 & 52.75 & 38.25 & 31.13 & 60.50 & 35.63 & 0.1562 & 0.8639 & 51.57 \\
		MiniCPM-o 2.6	  & 64.25 & 56.38 & 36.00 & 30.50 & 60.38 & 35.50 & 0.1280 & 0.9239 & 85.13 \\
		\midrule
		
		Qwen2-Audio~\cite{chu2024qwen2}        & 28.63 & 30.25 & 37.75 & 32.38 & 65.75 & 36.25 & 0.2717 & 0.7744 & 76.91 \\
		\textit{Qwen2-Audio-sft}  \\
		\quad EmoVoice-DB~\cite{3754829}   & 43.00 & 23.50 & 9.38 & 16.63 & 32.00 & 17.13 & 0.5351 & 0.4923 & 10.03 \\
		\quad TextrolSpeech~\cite{10445879} & 17.25 & 9.88 & 9.63 & 43.50 & 58.25 & 37.25 & 0.5691 & 0.3810 & 15.23 \\
		\quad \textbf{AffectSpeech (proposed)}  & 63.38 & 46.75 & 40.38 & \textbf{46.75} & \textbf{66.63} & 37.88 & 0.3569 & 0.7196 & 62.27 \\
		\midrule
		
		Qwen2.5-Omni~\cite{xu2025qwen2}       & 70.13 & 39.00 & 41.75 & 33.88 & 64.75 & 38.13 & 0.3190 & 0.7373 & 119.06 \\
		\textit{Qwen2.5-Omni-sft}  \\
		\quad EmoVoice-DB~\cite{3754829}   & 59.75 & 13.50 & 12.63 & 26.50 & 43.25 & 13.38 & 0.5038 & 0.5116 & 9.73 \\
		\quad TextrolSpeech~\cite{10445879} & 21.63 & 10.38 & 12.13 & 37.25 & 60.63 & 40.00 & 0.5667 & 0.3471 & 15.00 \\
		\quad \textbf{AffectSpeech (proposed)}  & \textbf{73.25} & 45.75 & \textbf{44.25} & \textbf{46.75} & 65.50 & \textbf{40.88} & 0.3998 & 0.6683 & 65.70 \\
		
		\bottomrule
	\end{tabular}
\end{table*}
\section{Speech Emotion Captioning on AffectSpeech}
\subsection{Experimental Setup}
To facilitate the learning of intricate speech-to-text mappings, we utilize 252,999 utterances from \textit{AffectSpeech}, paired with 1,517,994 multi-style descriptions for SFT. The remaining 800 utterances and their corresponding descriptions are reserved exclusively for evaluation, ensuring no overlapping samples between training and testing partitions.

We implement the SFT variants using Low-Rank Adaptation (LoRA)~\cite{hu2022lora} on Qwen2-Audio~\cite{chu2024qwen2} and Qwen2.5-Omni~\cite{xu2025qwen2} backbones. LoRA is applied to all linear layers with a rank $r=16$, $\alpha=32$, and a dropout rate of 0.05. Models are optimized for 3 epochs using the AdamW optimizer ($\beta_1=0.9, \beta_2=0.999$) with a weight decay of 0.05. We employ a peak learning rate of $1 \times 10^{-5}$ with a 5\% warmup phase and a cosine decay scheduler. All experiments are conducted on NVIDIA RTX PRO 6000 GPUs
with a batch size of 96.
A unified instruction template is used during inference to maintain consistency.

\subsection{Models for Comparison}
To benchmark the efficacy of \textit{AffectSpeech}, we fine-tune representative open-source SpeechLMs and compare them against a diverse range of state-of-the-art (SOTA) models on the same AffectSpeech test set to ensure a fair comparison. The comparison is organized into two groups:
\subsubsection{Foundation models (zero-shot)}
We assess the emotion understanding of proprietary models  (GPT-4o mini Audio, GPT-4o Audio), alongside open-source SpeechLMs and MLLMs, including Step-Audio 2 mini~\cite{wu2025step}, Audio Flamingo 3~\cite{ghosh2025audio}, and MiniCPM-o 2.6\footnote{\url{https://huggingface.co/openbmb/MiniCPM-o-2_6}}. These models are evaluated in a zero-shot setting to investigate whether general-purpose pre-training can capture the fine-grained affective nuances.
\subsubsection{Comparative SFT Variants}
To evaluate the impact of annotation granularity, we perform supervised fine-tuning on Qwen2-Audio~\cite{chu2024qwen2} and Qwen2.5-Omni~\cite{xu2025qwen2} using three distinct datasets: (i) EmoVoice-DB~\cite{3754829}, consisting of synthesized audios with sentence-level emotion descriptions; (ii) TextrolSpeech~\cite{10445879}, featuring coarse-grained prosodic captions; and (iii) our proposed \textit{AffectSpeech}, which provides large-scale speech with fine-grained, multi-style textual descriptions.

\subsection{Objective Evaluation}
\subsubsection{Emotion and prosody accuracy}
To ensure that the generated captions are physically anchored rather than being descriptive hallucinations, we quantify the classification accuracy of  emotion categories, intensity levels, and key segments to verify the understanding of speech emotions. Furthermore, we assess the model’s sensitivity to acoustic variations by measuring the precision of predicted prosodic attributes (pitch, tempo, and energy levels) against ground-truth metadata.
\subsubsection{Lexical diversity}
To quantify the model's capability to utilize a diverse emotional lexicon and avoid monotonous phrasing, we employ the Distinct-2~\cite{li2016diversity} metric by calculating the proportion of unique bigrams within the generated set as:
\begin{equation}
	\text{Distinct-}n = \frac{\text{Count}(\text{unique } n\text{-grams})}{\sum \text{Count}(n\text{-grams})},
\end{equation}
where $n=2$.
\subsubsection{Structural consistency}
To evaluate the structural regularity of descriptions, the 
Self-BLEU~\cite{zhu2018texygen} measure is utilized to measure intra-set redundancy by treating each generated caption $s_i$ as a hypothesis and the remaining set $S \setminus \{s_i\}$ as references.
\begin{equation}
	\text{Self-BLEU} = \frac{1}{N} \sum_{i=1}^{N} \text{BLEU}(s_i, S \setminus \{s_i\}),
\end{equation}
where $N$ denotes the total number of test utterances.

\subsection{Subjective Evaluation}
We conducted a human preference test to evaluate the perceptual alignment between utterances and generated captions, covering three dimensions: (i) emotion\footnote{Which caption more precisely identifies the emotion of the speech?}, (ii) prosody\footnote{Which caption better reflects the acoustic characteristics of the audio?}, and (iii) comprehension\footnote{Which caption provides the best analysis of the speech?}. 16 subjects (8 male, 8 female; age $25.9 \pm 2.6$) with normal hearing participated in a controlled environment. For each subject, 20 utterances were randomly 
sampled from the test set. The original audio was presented alongside three captions generated by models fine-tuned on EmoVoice-DB~\cite{3754829}, TextrolSpeech~\cite{10445879}, and the proposed \textit{AffectSpeech}, provided in a randomized, anonymous order.
Subjects selected the ``Best'' option or a ``Tie'' per dimension. Results are reported as the average preference rate, representing the percentage of times a specific SFT variant was favored.
\begin{figure}[htbp]
	\centering
	\includegraphics[width=1.0\columnwidth]{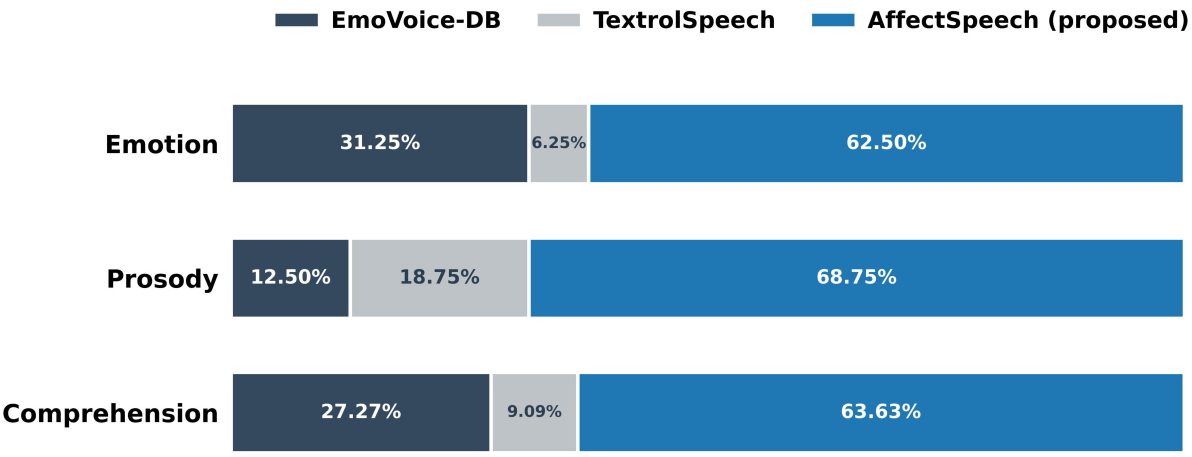}
	\caption{Subjective preference test results, which compare SFT variants trained on EmoVoice-DB, TextrolSpeech~\cite{10445879}, and the proposed AffectSpeech across three perceptual dimensions.}
	\label{fig-preference}
\end{figure}

\subsection{Results and Discussion}
\begin{table*}[t]
	\caption{Qualitative comparison of representative captions generated by different models and supervised Fine-tuning datasets.}
	\label{tab:caption_comparison}
	\centering
	\small
	\renewcommand{\arraystretch}{1.25} 
	\begin{tabularx}{\textwidth}{@{} l X @{}} 
		\toprule
		\textbf{Model} & \textbf{Output} \\ 
		\midrule
		Qwen2-Audio~\cite{chu2024qwen2} & The emotional tone is neutral. The speech emotion is calm. The emotional intensity is low. The pitch level is normal. the speech tempo is normal. the energy level is low. the key segment of emotional expression occurs in the entire part. \\ 
		\midrule
		Qwen2.5-Omni~\cite{xu2025qwen2} & The overall emotional tone is neutral; the speech emotion is calm with no particular intensity noted; the pitch level is normal; the speech tempo is normal; the energy level is moderate; the key segment of emotional expression occurs throughout the entire speech as there's no specific emphasis on any part. \\ 
		\midrule
		\textit{Qwen2.5-Omni-sft} & \\ 
		\quad EmoVoice-DB~\cite{3754829} & Expressing a neutral and balanced vocal quality. \\ 
		
		\quad TextrolSpeech~\cite{10445879} & The man speaks with a regular pitch, conversing at an average pace with standard energy. \\ 
		
		\quad \textbf{AffectSpeech (proposed)} & The audio has a neutral emotional tone and low emotional intensity. The speaker maintains a normal pitch, rate, and volume throughout, with no noticeable emotional fluctuations, conveying a calm and composed manner. \\ 
		\bottomrule
	\end{tabularx}
\end{table*}
The quantitative results of the SEC task are summarized in Table~\ref{tab:sec}, illustrating the performance of various state-of-the-art models and SFT variants. Across all metrics, models fine-tuned on our proposed \textit{AffectSpeech} dataset consistently outperform those fine-tuned on baseline datasets and demonstrate competitive or superior performance compared to proprietary zero-shot models. Notably, Qwen2.5-Omni fine-tuned on \textit{AffectSpeech} achieves the highest emotion category accuracy (73.25\%) and intensity precision (45.75\%), outperforming GPT-4o Audio by a 
large 
margin. This suggests that while large-scale general-purpose pre-training provides a robust foundation, specialized multi-granular affective annotations are indispensable for capturing subtle affective nuances.

Regarding emotion and prosody accuracy, the results highlight a distinct advantage for the \textit{AffectSpeech} SFT variants. While the models fine-tuned on EmoVoice-DB~\cite{3754829} shows reasonable category accuracy, they fail to generalize to intensity and key segment detection due to their coarse-grained characteristics. Conversely, TextrolSpeech~\cite{10445879} variants exhibit higher prosody accuracy in pitch and tempo but struggle with emotion classification. \textit{AffectSpeech} effectively bridges this gap; for instance, the Qwen2.5-Omni SFT variant achieves 46.75\% pitch accuracy and 65.50\% tempo accuracy, 
considerably 
surpassing its zero-shot counterpart and establishing  more reliable perception.

The analysis of description quality, including Distinct-2~\cite{li2016diversity}, Self-BLEU~\cite{zhu2018texygen}, and average word count, reveals a trade-off between lexical diversity and structural regularity. Unlike traditional metrics where higher or lower values are often unilaterally preferred, SEC requires a delicate balance to ensure both stylistic consistency and descriptive richness. Baselines such as EmoVoice-DB~\cite{3754829} and TextrolSpeech~\cite{10445879} exhibit high Distinct-2 scores and low Self-BLEU, but this is primarily a byproduct of their extremely short average word counts ($\approx 10$--$15$ words), which lack the complexity required for comprehensive analysis. In contrast, zero-shot models like Step-Audio 2 mini~\cite{wu2025step} and Audio Flamingo 3~\cite{ghosh2025audio} show high Self-BLEU and low Distinct-2 values, indicating a tendency toward repetitive descriptions. Our \textit{AffectSpeech} variants occupy a ``sweet spot,'' maintaining moderate Distinct-2 ($\approx 0.4$) and Self-BLEU ($\approx 0.67$) value with an informative average length of approximately 65 words. This combination indicates that the model has successfully internalized the structured stylistic paradigms of our dataset while retaining sufficient lexical diversity to describe unique emotional instances vividly.

Subjective preference results, as illustrated in Fig.~\ref{fig-preference}, corroborate the objective evaluations, with the proposed \textit{AffectSpeech} SFT variant consistently emerging as the most favored model across all dimensions.
In terms of emotional description, \textit{AffectSpeech} achieved a dominant preference rate of 62.50\%, demonstrating that our multi-granular, open-vocabulary descriptions offer the necessary depth for human perception of high-level affective intent.
The margin is even more pronounced in the prosody dimension (68.75\%), which suggests that our dataset's alignment with acoustic characteristics enables models to articulate nuanced prosodic variations that coarse-grained alternatives fail to capture.
For overall comprehension, \textit{AffectSpeech} maintained its lead, reflecting its superior ability to synthesize complex affective information into a coherent description. These subjective results confirm that the structural and lexical richness of \textit{AffectSpeech} translates into captions that are perceived as 
strongly 
more ``grounded'' and perceptually accurate by human listeners.

To further illustrate these differences, qualitative examples are compared in Table~\ref{tab:caption_comparison}. While baseline models often produce overly simplistic outputs, the proposed \textit{AffectSpeech} SFT variant exhibits a more refined and comprehensive articulation of both acoustic and affective nuances.

\begin{table*}[ht]
	\centering
	\footnotesize
	\setlength{\tabcolsep}{4.6pt}
	\caption{Quantitative and Qualitative Comparisons of Emotional Speech Synthesis with Textual Description.}
	\label{tab:ess}
	
	\begin{tabular}{
			l
			ccc
			cccc
		}
		\toprule
		\multirow{2}{*}{\textbf{Model}} &
		\multicolumn{4}{c}{\textbf{Objective Evaluation}} &
		\multicolumn{3}{c}{\textbf{Subjective Evaluation}} \\
		
		\cmidrule(lr){2-5}
		\cmidrule(lr){6-8}
		
		& \textbf{WER [\%] $\downarrow$} & \textbf{DDUR [s] $\downarrow$} & \textbf{Emo\_Sim [\%] $\uparrow$}
		& \textbf{Emo\_Div [$\times 10^{-4}$] $\uparrow$} & \textbf{Audio Quality}
		& \textbf{Intelligibility} & \textbf{Naturalness} \\
		\midrule		
		GPT-4o mini TTS  & 4.72 & 0.27 & 66.85 & 15.69  & 3.97$\pm$0.15 & 4.03$\pm$0.21 & 4.08$\pm$0.25 \\
		TTS-1     & 4.28 & 0.23 & 85.95 & 14.51  & 3.59$\pm$0.15 & 3.79$\pm$0.20 & 3.62$\pm$0.21 \\
		TTS-1 HD  & \textbf{1.11} & \textbf{0.21} & 80.07 & 14.07  & 3.72$\pm$0.26 & 3.83$\pm$0.21 & 3.31$\pm$0.26 \\
		PromptTTS~\cite{guo2023prompttts}  & 15.73 & 0.34 & 64.13 & 6.85  & 2.97$\pm$0.17 & 2.95$\pm$0.12 & 3.15$\pm$0.23 \\
		Salle~\cite{10445879}      & 10.97 & 0.23 & 70.14 & 13.75  & 3.49$\pm$0.23 & 3.36$\pm$0.28 & 3.22$\pm$0.27 \\
		\midrule
		
		VoxInstruct~\cite{zhou2024voxinstruct}      & 7.40 & 0.27 & 83.23 & 19.23  & 3.73$\pm$0.22 & 3.80$\pm$0.21 & 3.59$\pm$0.18 \\
		\textit{VoxInstruct-sft}  \\
		\quad EmoVoice-DB~\cite{3754829}   & 4.52 & 0.26 & 86.14 & 17.35  & 3.55$\pm$0.26 & 3.72$\pm$0.19 & 3.46$\pm$0.17 \\
		\quad TextrolSpeech~\cite{10445879} & 4.73 & 0.29 & 82.75 & 19.40  & 3.58$\pm$0.24 & 3.59$\pm$0.21 & 3.66$\pm$0.20 \\
		\quad \textbf{AffectSpeech (proposed)}  & 4.36 & 0.28 & \textbf{86.40} & \textbf{19.91} & 3.76$\pm$0.20 & 3.94$\pm$0.23 & 3.81$\pm$0.25 \\
		\midrule
		
		Ground Truth        & 0.56 & - & - & 22.04  & 4.13$\pm$0.15 & 4.40$\pm$0.16 & 4.31$\pm$0.22 \\
		\bottomrule
	\end{tabular}
\end{table*}
\section{Emotional Speech Synthesis on AffectSpeech}
\subsection{Experimental Setup}
The data partitionings for the ESS task are consistent with those described in Section IV.A, utilizing the same 252,999 SFT utterances and a held-out evaluation set of 800 samples.
Unlike SEC, ESS additionally requires explicit specification of the textual content, where both the emotional description is provided as the instruction and the linguistic content is given as the target text to be synthesized.

We implement the SFT variants using VoxInstruct~\cite{zhou2024voxinstruct} as the backbone. To adapt the model for high-fidelity emotional synthesis while maintaining linguistic stability, LoRA~\cite{hu2022lora} is applied to all linear layers with a rank $r=8$ and $\alpha=16$. The optimization strategy follows the SEC setup, utilizing the AdamW optimizer with a $1 \times 10^{-5}$ peak learning rate and a cosine scheduler over 3 epochs. All experiments are conducted on NVIDIA RTX PRO 6000 GPUs
with a batch size of 96.

\subsection{Models for Comparison}
To validate the synthesis performance enabled by \textit{AffectSpeech}, we fine-tune the representative open-source SpeechLM
and compare against two groups of SOTA models:
\subsubsection{Foundation models (zero-shot)}
We evaluate the emotional synthesis capabilities of proprietary models, including GPT-4o mini TTS, OpenAI TTS-1, and TTS-1 HD. Additionally, we include representative open-source ESS frameworks: PromptTTS~\cite{guo2023prompttts}, and Salle~\cite{10445879}.
\subsubsection{Comparative SFT Variants}
To isolate the impact of textual granularity on generative control, we fine-tune the VoxInstruct~\cite{zhou2024voxinstruct} backbone on three corpora: (i) EmoVoice-DB~\cite{3754829}, (ii) TextrolSpeech~\cite{10445879}, and (iii) our proposed \textit{AffectSpeech}. By fixing the model architecture, we directly measure how fine-grained descriptions influence the nuance and prosodic diversity of the synthesized speech.

\subsection{Objective Evaluation}
\subsubsection{Linguistic integrity}
To measure the precision of pronounciation, we calculate the word error rate (WER) by measuring transcription errors using whisper-large-v3\footnote{https://huggingface.co/openai/whisper-large-v3} as:
\begin{equation}
	\text{WER} = \frac{S_{\text{word}} + D_{\text{word}} + I_{\text{word}}}{N_{\text{word}}},
\end{equation}
where $S_{\text{word}}$ denotes the number of words that are replaced by a different word in the predicted output, $D_{\text{word}}$ denotes the number of words that are missed, $I_{\text{word}}$ denotes the number of extra words in the predicted output that are not present in the speech, and $N_{\text{word}}$ denotes the total number of words.

\subsubsection{Rhythmic similarity}
To evaluate the rhythm of synthesized speech, we compute the average duration difference (DDUR) between the voiced segments of synthesized and ground-truth utterances as:
\begin{equation}
	\text{DDUR} = \frac{1}{N} \sum_{i=1}^{N} |dur_i - \hat{dur}_i|,
\end{equation}
where $dur_i$ denotes the duration of the $i$-th ground-truth sample, and $\hat{dur}_i$ denotes the duration of the $i$-th synthesized sample.

\subsubsection{Emotion similarity}
The cosine similarity between the synthesized and ground-truth utterances is calculated to quantifies the faithfulness of the generated speech to its textual description. 
Specifically, we extract emotion embeddings $\mathbf{\hat{e}}_i$ and $\mathbf{e}_i$ from the synthesized and ground-truth speech, respectively, using the pre-trained Emotion2Vec~\cite{ma2024emotion2vec}. The average emotion similarity is defined as:
\begin{equation}
	\text{Emo\_Sim} = \frac{1}{N} \sum_{i=1}^{N} \frac{\mathbf{e}_{i} \mathbf{\hat{e}}_{i}}{\|\mathbf{e}_{i}\|_2 \|\mathbf{\hat{e}}_{i}\|_2},
\end{equation}
where $\mathbf{e}_i$ represent the emotion embeddings of the $i$-th sample.

\subsubsection{Emotion diversity}
The average pairwise distance is employed to quantitatively evaluate the diversity and coverage of the emotional expressions. Specifically, we extract emotion embeddings $\{\mathbf{e}_i\}_{i=1}^N$ from the speech using Emotion2Vec~\cite{ma2024emotion2vec}, and calculate the mean Euclidean distance as:
\begin{equation}
	\text{Emo\_Div} = \frac{2}{N(N-1)} \sum_{i=1}^{N-1} \sum_{j=i+1}^{N} \| \mathbf{e}_i - \mathbf{e}_j \|_2.
\end{equation}

\subsection{Subjective Evaluation}
To assess the perceptual quality and emotional expressiveness of the synthesized speech, we conducted two listening tests involving the same 16 subjects described in Section IV.D.
For each evaluator, 30 utterances were randomly sampled from the test set, resulting in 330 evaluation trials.
\subsubsection{Mean opinion score (MOS)}
A standard MOS test was conducted to evaluate the general synthesis quality across audio quality, intelligibility, and naturalness. Each aspect is rated on a five-point scale: $1~\text{(very poor)} \rightarrow 5~\text{(excellent)}$.
Ground-truth utterances were included as references to provide an upper-bound benchmark for performance. All MOS scores are averaged across participants.
\subsubsection{Perceptual attribute analysis}
To further investigate the impact of fine-grained textual descriptions on the controllability of ESS, we performed a specialized listening test based on seven-point Likert scale~\cite{joshi2015likert}: $1~\text{(strongly disagree)} \rightarrow 7~\text{(strongly agree)}$.
Participants were asked to evaluate the degree to which the synthesized samples satisfied three orthogonal dimensions including prosodic fluency, emotion accuracy, and emotion diversity. 
\begin{figure}[tbp]
	\centering
	\includegraphics[width=1.0\columnwidth]{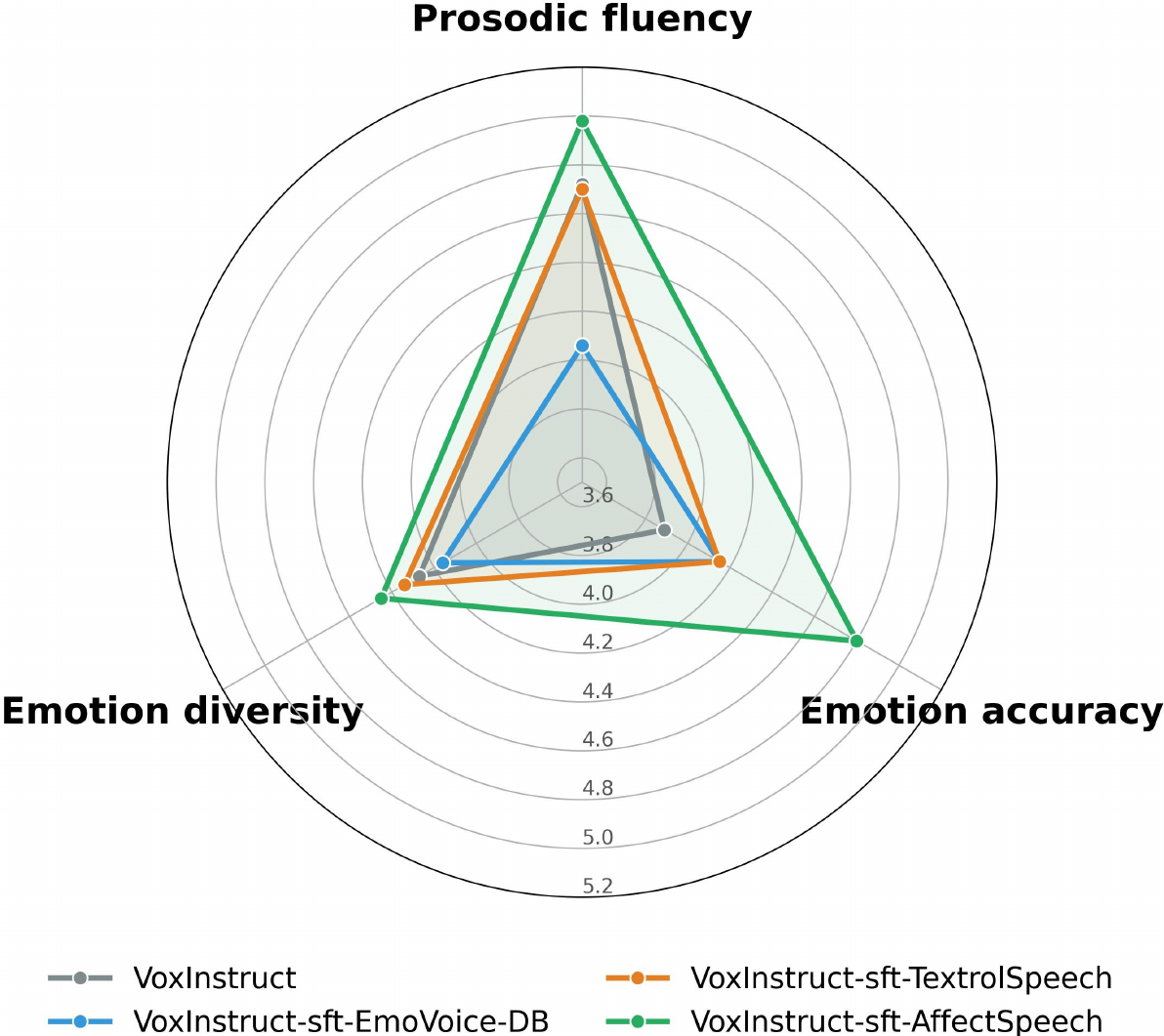}
	\caption{Perceptual attribute analysis of synthesized speech, illustrating subjective ratings (1--7 Likert scale) for prosodic fluency, emotion accuracy, and emotion diversity.}
	\label{fig-radar}
\end{figure}
\subsection{Results and Discussion}
The quantitative and qualitative performance of ESS is shown in Table \ref{tab:ess}, with perceptual attribute analysis further illustrated via the radar chart in Fig. \ref{fig-radar}.

Objective results demonstrate that the proposed \textit{AffectSpeech} dataset 
considerably 
enhances the synthesis capabilities of the VoxInstruct~\cite{zhou2024voxinstruct} backbone.
Specifically, the VoxInstruct fine-tuned on \textit{AffectSpeech} achieves a WER of 4.36\%, which is competitive with the proprietary models such as  OpenAI's TTS-1 and GPT-4o mini TTS.
This suggests that the large-scale, fine-grained descriptions in \textit{AffectSpeech} do not compromise linguistic integrity; rather, they provide sufficient stylistic guidance to maintain stable pronunciation while allowing for complex emotional rendering.
In terms of emotional similarity and diversity, the proposed \textit{AffectSpeech} SFT variant achieves the highest Emo\_Sim and Emo\_Div among all models. While EmoVoice-DB~\cite{3754829} shows comparable emotion similarity (86.14\%), its emotion diversity is noticeably lower, likely due to the limited-scale synthesized speech that also lacks expressive authenticity. In contrast, the large-scale, human-recorded speech with fine-grained annotations in \textit{AffectSpeech} empower the model to explore a wider range of the emotional representation space.

The subjective MOS test further validates these objective results. The \textit{AffectSpeech} variant consistently outperforms other SFT baselines in audio quality, intelligibility, and naturalness. Notably, it 
clearly 
surpasses PromptTTS~\cite{guo2023prompttts} and Salle~\cite{10445879}, approaching the performance of proprietary foundation models especially for GPT-4o mini TTS. The results indicate that the fine-grained information provided by multi-granular descriptions helps the model avoid the flat prosody common in coarse-grained ESS, resulting in speech that listeners perceived as more realistic and better aligned with the provided textual descriptions.
Perceptual attribute analysis, as visualized in Fig.\ \ref{fig-radar}, provides more insights into the impact of description granularity on synthesis control. The \textit{AffectSpeech} SFT variant model forms the outermost contour, dominating in prosodic fluency, emotion accuracy, and emotion diversity. Compared to the original VoxInstruct~\cite{zhou2024voxinstruct}, fine-tuning on \textit{AffectSpeech} 
strongly 
expands the model's capacity to synthesize varied emotional states accurately. While the TextrolSpeech~\cite{10445879} variant shows reasonable prosodic fluency, it lacks the emotion accuracy provided by our dataset's detailed emotional categorization. Conversely, EmoVoice-DB~\cite{3754829} achieves moderate accuracy but fails to match the diversity and fluency enabled by \textit{AffectSpeech}. This  superiority confirms that the integration of multidimensional characteristics in our descriptions allows the model to map textual description to emotional prosody more effectively.

\section{Conclusion}
In this paper, we introduced \textit{AffectSpeech}, a large-scale emotional speech dataset enriched with fine-grained textual descriptions to overcome the limitations of traditional categorical and low-dimensional emotion representations. By shifting the paradigm toward natural language descriptions, \textit{AffectSpeech} provides a more flexible and interpretable framework for modeling the complex nuances of human vocal expression.

To address the challenges of annotation scalability and reliability, we established a novel human--LLM collaborative pipeline. This approach effectively integrates algorithmic pre-labeling, multi-LLM generation, and  human-in-the-loop verification to curate high-quality annotations across six complementary dimensions, including sentiment polarity, intensity, prosodic attributes, and prominent segments. Furthermore, by reformulating these annotations into diverse functional styles, we 
effectively 
enhanced the linguistic diversity of the corpus, mitigating the risk of stylistic overfitting in foundation models.

Experimental results on SEC and ESS demonstrate the superiority of our dataset. Models fine-tuned on \textit{AffectSpeech} consistently outperformed baselines, validating that fine-grained, multi-dimensional textual descriptions are crucial for capturing the intricate interplay between semantics, prosody, and emotion.
In future work, we plan to extend \textit{AffectSpeech} to a multilingual dataset while further enriching with finer-grained and more comprehensive emotion annotations.

\small
\bibliographystyle{IEEEtran}
\bibliography{references}

\begin{IEEEbiography}
	[{\includegraphics[width=1in,height=1.25in,clip,keepaspectratio]{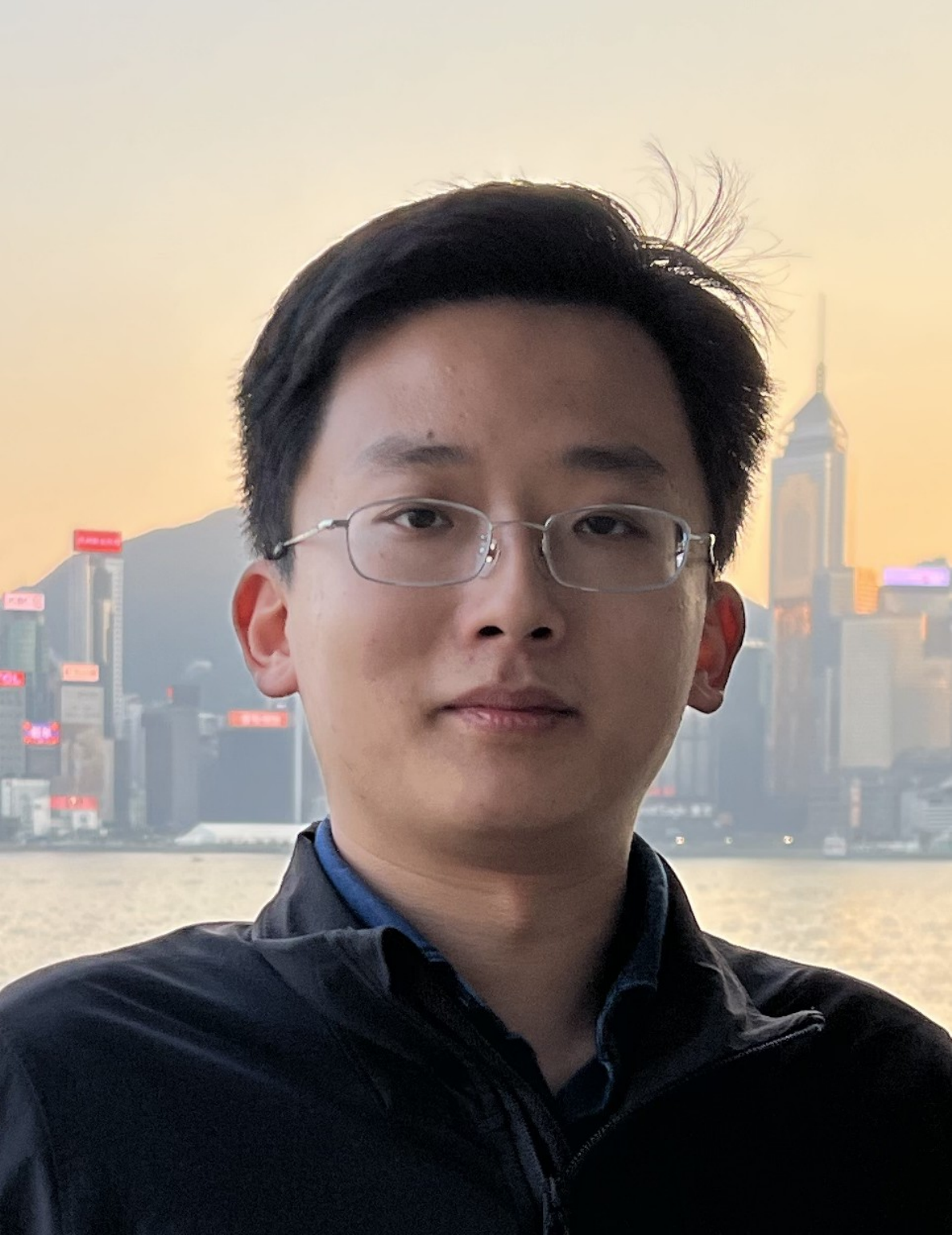}}]{Tianhua Qi}
	(Student Member, IEEE) is currently working toward the PhD degree with the School of Biological Science and Medical Engineering, Southeast University, Nanjing 210096, China, and also with the Key Laboratory of Child Development and Learning Science (Southeast University), Ministry of Education, China. He is the general co-chair of The 12th ISCA-SAC Doctoral Consortium. His research interests include affective computing, deep learning, and speech signal processing.
\end{IEEEbiography}

\begin{IEEEbiography}[{\includegraphics[width=1.0in,height=1.25in,clip,keepaspectratio]{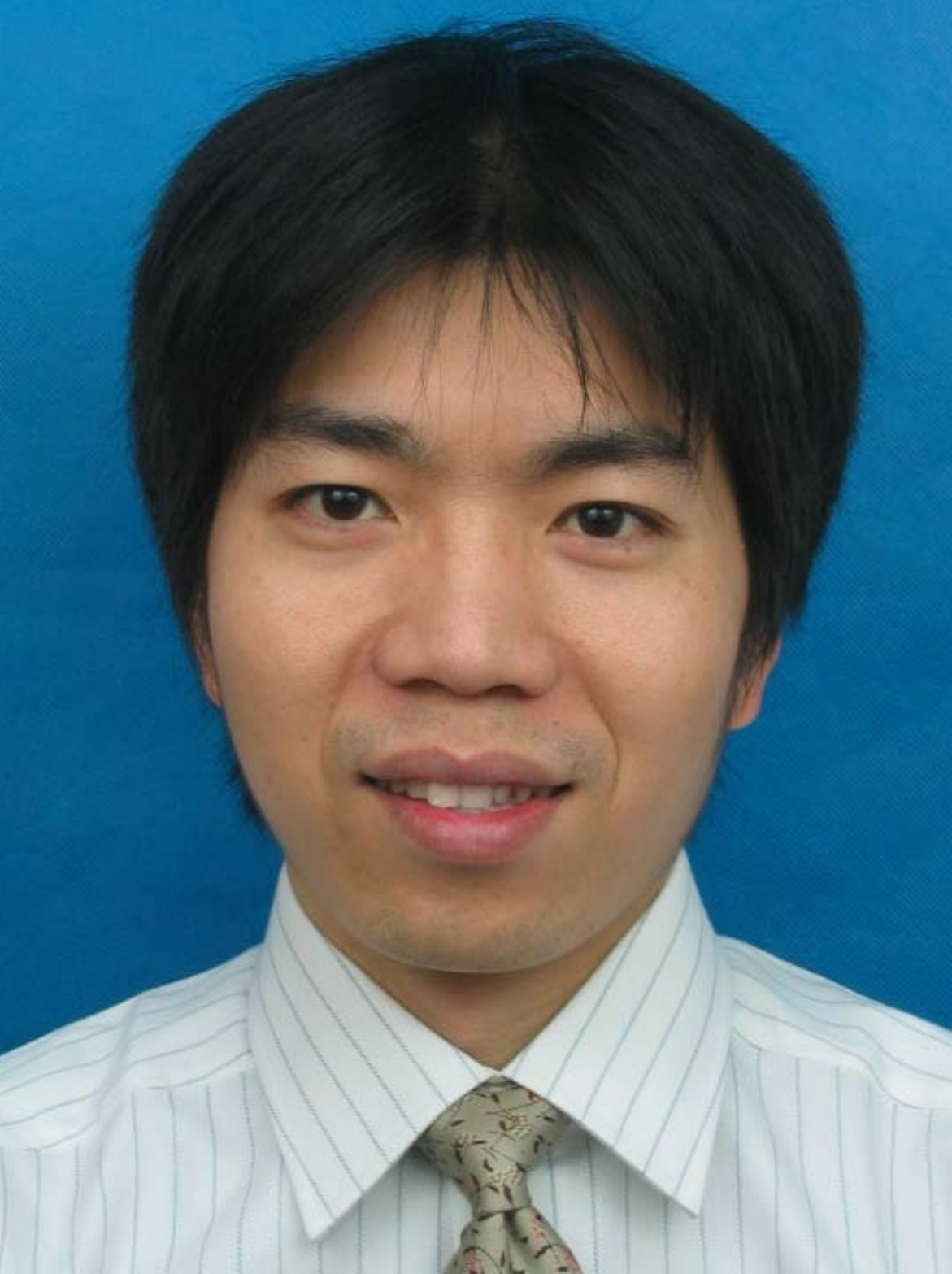}}]{Wenming Zheng} 
	(Senior Member, IEEE) received the B.S. degree in computer science from Fuzhou University, Fuzhou, China, in 1997, the M.S. degree in computer science from Huaqiao University, Quanzhou, China, in 2001, and the Ph.D. degree in signal processing from Southeast University, Nanjing, China, in 2004. Since 2004, he has been with the Research Center for Learning Science, Southeast University.	Currently, he is a Professor with the Key Laboratory of Child Development and Learning Science (Ministry of Education), School of Biological Sciences and Medical Engineering, Southeast University. 
	He has been elected as Institution of Engineering and Technology (IET) Fellow since 2022. 
	His research interests include affective computing, pattern recognition, machine learning, and computer vision. He is a former associate editor of the IEEE Transactions on Affective Computing, the associate editor of the IEEE Transactions on Cognitive and Developmental Systems, and the editorial board member of The Visual Computer.
\end{IEEEbiography}

\begin{IEEEbiography}[{\includegraphics[width=1.0in,height=1.25in,clip,keepaspectratio]{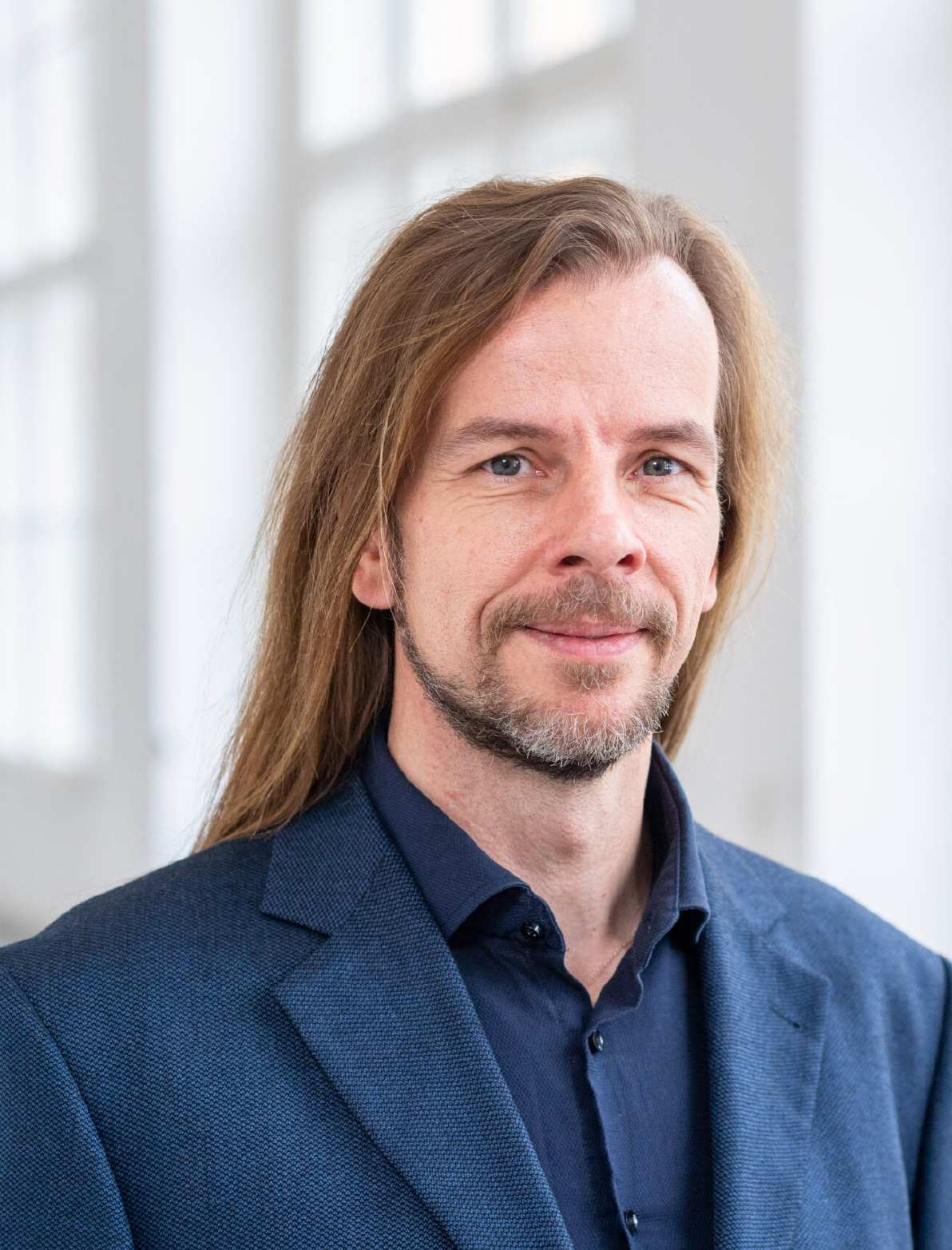}}]{Bj{\"o}rn W. Schuller} 
	(Fellow, IEEE) received the diploma, in 1999, the doctoral degree, in 2006, and	the habilitation and adjunct teaching professorship in signal processing and machine intelligence, in 2012, all in electrical engineering and information technology from Technische Universit{\"a}t M{\"u}nchen (TUM), Germany. He is a tenured full professor heading CHI -- the Chair of Health Informatics at the TUM University Hospital, and a professor of Artificial Intelligence heading GLAM -- the Group on Language, Audio, \& Music, with the Imperial College London in London, U.K. He is the field chief editor of Frontiers in Digital Health, editor-in-chief of the AI Open Journal, former editor in chief of the IEEE Transactions on Affective Computing, President-emeritus and Fellow of the AAAC, Golden Core Awardee of the IEEE Computer Society, Fellow of the ACM, Fellow of the ISCA, and Fellow of the ELLIS and BCS. He (co-)authored 5 books and more than 1,700 publications in peer reviewed books, journals, and conference proceedings leading to 80 k citations (h-index 124).
\end{IEEEbiography}

\begin{IEEEbiography}[{\includegraphics[width=1.0in,height=1.25in,clip,keepaspectratio]{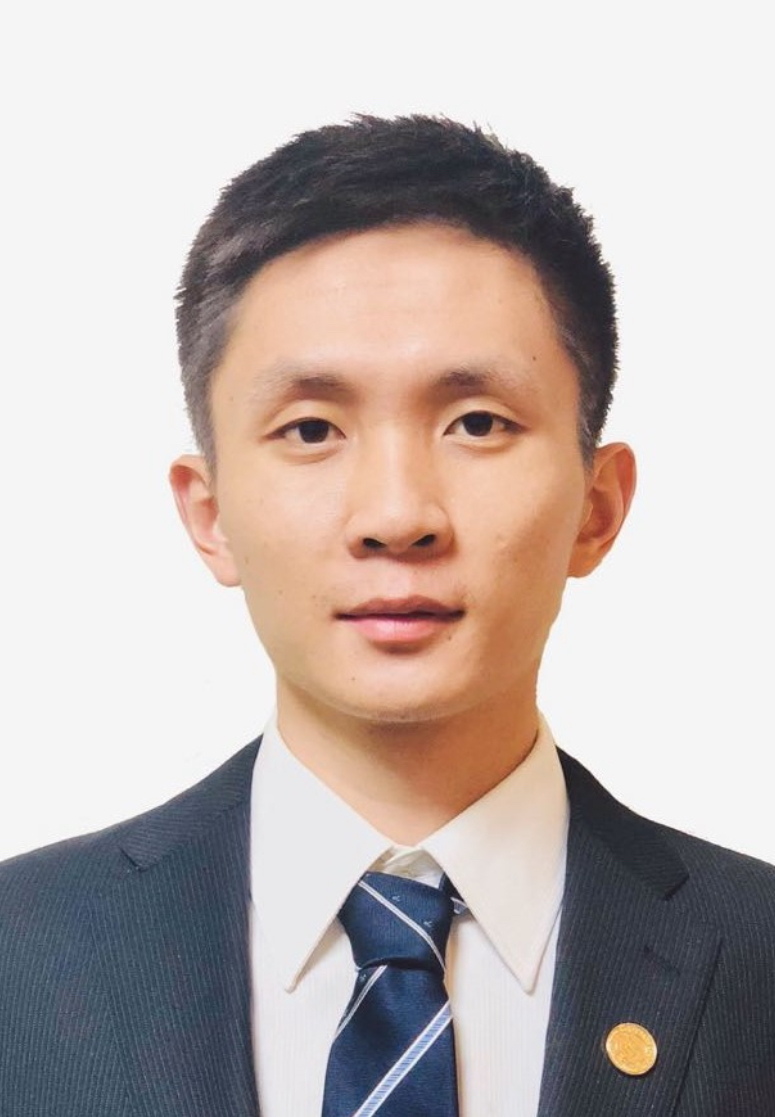}}]{Zhaojie Luo} 
	(Member, IEEE) received the M.Eng. and Dr.Eng. degrees from Kobe University, Kobe, Japan, in 2017 and 2020, respectively. He is currently an Associate Professor with Southeast University, Nanjing, China. From 2020 to 2024 he was an assistant Professor with Osaka University, Suita, Japan. From 2019 to 2020, he was a Researcher with the Department of Electrical \& Computer Engineering, National University of Singapore, Singapore. His research interests include voice conversion, speech synthesis, facial expression recognition, multimodal emotion recognition, and statistical signal processing. He has published more than 20 papers in top-tier speech/multimedia journals and international conferences, such as IEEE/ACM Transactions on Audio, Speech and Language Processing, IEEE Transactions on Multimedia, EURASIP JASMP, ACM Multimedia, INTERSPEECH, ICASSP, SSW, ICME, and ICPR. He is a member of ISCA and ASJ, and is the reviewer for many major referred journal and conference papers.
\end{IEEEbiography}

\begin{IEEEbiography}[{\includegraphics[width=1.0in,height=1.25in,clip,keepaspectratio]{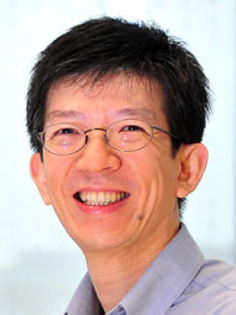}}]{Haizhou Li} 
	(Fellow, IEEE) received the B. Sc., M. Sc., and Ph.D degree in electrical and electronic engineering from South China University of Technology, Guangzhou, China in 1984, 1987, and 1990 respectively. He is currently a Presidential Chair Professor and the Executive Dean of the School of Data Science, The Chinese University of Hong Kong, China. He is also an Adjunct Professor with the Department of Electrical and Computer Engineering, National University of Singapore, Singapore. His research interests include automatic speech recognition, speaker and language recognition, natural language processing. He was the Editor-in-Chief of IEEE/ACM Transactions on Audio, Speech and Language Processing during 2015–2018, an elected Member of IEEE Speech and Language Processing Technical Committee during 2013–2015, the President of the International Speech Communication Association during 2015–2017, President of Asia Pacific Signal and Information Processing Association during 2015–2016, and President of Asian Federation of Natural Language Processing during 2017–2018. Since 2012, he has been a Member of the Editorial Board of Computer Speech and Language. He was the General Chair of ACL 2012, INTERSPEECH 2014, ASRU 2019 and ICASSP 2022. He is a Fellow of the ISCA, and a Fellow of the Academy of Engineering Singapore. He was the recipient of the National Infocomm Award 2002, and President’s Technology Award 2013 in Singapore. He was named one of the two Nokia Visiting Professors in 2009 by the Nokia Foundation, and U Bremen Excellence Chair Professor in 2019.
\end{IEEEbiography}

\end{document}